\documentclass[aps, prd, nofootinbib]{revtex4}
\usepackage{amsmath}
\usepackage{graphicx}
\usepackage{dcolumn}
\usepackage{bm}
\usepackage{amssymb}
\usepackage{latexsym}
\usepackage{bm}
\usepackage{mathrsfs}
\usepackage{booktabs}
\usepackage{float}

\newcommand{\no}{\nonumber}

\newcommand{\pa}{\partial}

{\bf }

\newcommand{\half}{\frac{1}{2}}

\begin{document}


\title{Understanding  quantum behaviors of an electron in a uniform magnetic field alternatively}

\vskip 2mm


\author{Jin-Ming Wang}

\author{Yuan-Zao Gao}

\author{Dai-Lin Cun}

\author{Jian Jing }
\email{jingjian@mail. buct. edu. cn}



\affiliation{{Department of Physics and Electronics, School of Science, Beijing University of Chemical Technology, Beijing 100029, P. R. China.}}



\begin{abstract}
Quantum mechanically, an electron moving in a uniform magnetic field forms Landau levels. A curious feature is that for states with a negative angular quantum number, the total probability current vanishes, which appears to contradict the classical picture of cyclotron motion. While a geometric interpretation based on classical orbits exists, alternative interpretations remain of interest. In this paper, we examine the probability current density and identify a critical radius that naturally partitions the plane into an inner clockwise-flow region and an outer counterclockwise-flow region. We show that the vanishing total current results from an exact cancellation between these two regions. Furthermore, by defining a partitioned kinetic angular momentum with respect to the critical radius, we reveal an intrinsic competitive structure: the electron simultaneously carries two opposing rotational components. The negative quantum number manifests in the strength of the inner counter-rotation, while the net kinetic angular momentum remains positive. This bidirectional flow picture also provides a dynamical interpretation of the infinite degeneracy of Landau levels.

\end{abstract}




\maketitle

\section{Introduction}

The Landau level problem, whose classical counterpart describes electron motion in a uniform magnetic field, was exactly solved in the 1930s \cite{Landau, Landautb}. It provides  the theoretical groundwork for the quantum Hall effect  \cite{QHE1, QHE2, QHE3} and remains a central topic in fundamental physics research \cite{JL, NGM, SSSB, GSFB, GSFB2},  particularly in the studies of the quantum-classical correspondence \cite{Page, GFS, Jap1, Jap2, Jap3, Jap4}.

It is known that in the symmetric gauge,  
the canonical angular momentum commutes with the Hamiltonian, thus the two operators share a common set of eigenstates. These eigenstates are labeled by the radial quantum number $n_r$ and  the angular quantum number $m$. The probability current density associated with eigenstates $|n_r, m \rangle$ and its relation to classical cyclotron motion have been examined \cite{Capri}. A particularly noteworthy observation is the following: for $m>0$, the total probability current is consistent with the classical picture; for $m=0$, the total current is exactly half of the classical value; and for $m<0$ the total current vanishes identically. To account for this phenomenon, an illuminating geometric interpretation has been proposed in the literature, which compares the distance from the orbit center to the origin with the cyclotron radius, focusing on whether the classical orbit encloses the coordinate origin \cite{Li}.

Although this geometric picture  is instructive, 
it is still of interest  to examine its physical origin from a different perspective. In this paper, we analyze this phenomenon from the perspective of the probability current density. It is observed that the probability current density can be decomposed into two parts, one part is related to the canonical  momentum, the other part is related to the gauge field. For eigenstates with $m < 0$, these two contributions have opposite signs. Their competition gives rise to a critical radius. Starting from this  structure, we address two interrelated questions. One is the physical origin of the vanishing total current for negative $m$ Landau eigenstates, the other is the way the negative quantum number $m$ manifest itself in the  kinetic  angular momentum. 

Recently, the study of quantum Hall states has been extended to curved surfaces, particularly in the context of three-dimensional topological insulator nanowires. In Refs. \cite{han1, han2}, the authors show  that Dirac quantum Hall states on axially symmetric curved surfaces exhibit a geometric branch whose energy scaling with the magnetic field is sensitive to the surface shape. Moreover, for surfaces with constant negative curvature, such as the pseudosphere and the Minding surface, analytical and numerical solutions of the Dirac equation reveal that a coaxial magnetic field gives rise to a peculiar $B^{{1}/{4}}$  scaling for one class of Landau levels, in contrast to the usual $B^{{1}/{2}}$   behavior in the planar case. These findings highlight the rich interplay between geometry, magnetic field, and quantum flow. In this context, our present analysis of the bidirectional flow structure in the planar Landau problem may offer a complementary perspective on the internal current patterns that also underlie such curved-space quantum Hall systems.

The organization of this paper is as follows: In the next section, after reviewing the classical aspect of the model, we pay our attention to its quantum aspect. We  show that for the negative angular momentum quantum number, there is a critical radius. The probability current density points in opposite directions on either side of this critical radius. In {\bf section III}, we provide an alternative interpretation to the problem of the  vanishing total current for the negative $m$  eigenstates. Then, we study the role the negative $m$ eigenstates played in the kinetic angular momentum   in {\bf section IV}.  Some remarks and conclusions are given in the last section.

\section{The model and the critical radius}

Classically, an electron moves in uniform circular motion in a plane perpendicular to a uniform magnetic field. The relation between its speed and orbital radius is
\begin{equation}
v = \frac{|e|B R}{\mu} \label{cv}
\end{equation}
where $B, R,  \ e =-|e| \ {\rm and }\ \mu$ are  magnitude of the magnetic field, radius of the orbit, the charge and mass of the electron.  From Eq. (\ref{cv}), it is easy to  derive the classical expression for the electric current
\begin{equation}
i_c = \frac{v}{2 \pi R} = \frac{|e|B}{{2 \pi \mu}} \label{cp}
\end{equation}
and the relation between the kinetic energy $E$ and kinetic angular momentum $L$
\begin{equation}
E = \frac{|e|B}{2 \mu} L. \label{cel}
\end{equation}

The quantum dynamics of this electron is governed by the  Hamiltonian
\begin{equation}
H = \frac{\bm \pi^2}{2 \mu} \label{ha1}
\end{equation}
where $\bm \pi$ is the kinetic  momentum, which relates to the canonical momentum  $\mathbf p$ via
\begin{equation}
\bm \pi=  \mathbf p + e \mathbf A
\end{equation}
with
$\mathbf A$ being the gauge field. In this paper, we choose the symmetric gauge
$\mathbf A = \half \mathbf B \times \mathbf r$,
where $\mathbf B$ is the uniform magnetic field along the $z$-axis, i.e., $ \mathbf B = B \mathbf {e_z} $. In polar coordinate, the symmetric gauge takes the form
\begin{equation}
\mathbf A = \half B r \mathbf e_\phi \label{sg1}
\end{equation}
where $\mathbf e_\phi$ is the unit vector along the azimuthal direction.


Choosing symmetric gauge (\ref{sg1}) and neglecting the free motion along the $z$-axis, we write the Hamiltonian (\ref{ha1}) and canonical angular momentum in the polar coordinate $(r, \ \phi)$ as
\begin{equation}
H= - \frac{\hbar^2}{2 \mu r} \frac{\pa }{\pa r} \Big(r \frac{\pa}{\pa r} \Big)  - \frac{\hbar^2}{2 \mu} \Big( \frac{\pa}{r \pa \phi} - \frac{i e B r}{2 \hbar} \Big)^2 \label{ha3}
\end{equation}
and
\begin{equation}
L_z = - i \hbar \frac{\pa}{\pa \phi}.
\end{equation}
It can be checked directly that
\begin{equation}
[L_z, \ H]=0. \label{camh}
\end{equation}
It means that the canonical angular momentum and the Hamiltonian shares a  common set of eigenstates  $|n_r, m \rangle$.

The eigenvalues of the Hamiltonian (\ref{ha3}) are 
\begin{eqnarray}
E_{n,m} &&= \Big (n_r + \frac{|m| + m +1}{2} \Big) \frac{\hbar |e| B}{\mu}, \label{evs} \\
&& n_r =0,1,2, \cdots, \ m =0, \pm 1, \pm2, \cdots.  \no
\end{eqnarray}
The normalized common eigenstates are
\begin{equation}
\psi_{n_r, m} (r, \phi) = \langle r, \phi|n_r, m\rangle = \frac{1}{\sqrt {2 \pi}} R_{n_r,m} e ^{im \phi} \label{nes}
\end{equation}
where $R_{n_r,m}$ is the radial wavefunction,
\begin{eqnarray}
R_{n_r,m} =&& \frac{1}{a^{1+|m|}} \Big (\frac{n_r !}{2 ^{|m|}(n_r + |m|)!} \Big)^\half  r^{|m|} \no \\
&&\times \exp (-\frac{r^2}{4a^2})L_{n_r}^{|m|} (\frac{r^2}{2 a^2})
\end{eqnarray}
in which  $a=(\frac{\hbar}{|e| B})^ \half$ is  the  magnetic length and $L_{n_r}^{|m|}$ is the associated Laguerre polynomial.

The probability current density in the eigenstates $\psi_{n_r, m}$ is
\begin{eqnarray}
\mathbf J &=& \mathbf J^{\rm can} + \mathbf J^{\rm gau} \no \\
&=& \frac{1}{2 \mu}(\psi^* _{n_r, m} \mathbf p \psi_{n_r, m} - \psi_{n_r, m} \mathbf p \psi^* _{n_r, m}) \no \\
&&- \frac{e}{\mu} \mathbf A \psi^* _{n_r,m} \psi _{n_r,m} \label{pcd}
\end{eqnarray}
in which
\begin{eqnarray}
\mathbf J^{\rm can}&=& \frac{1}{2 \mu}(\psi^* _{n_r,m} \mathbf p \psi_{n_r,m} - \psi_{n_r,m} \mathbf p \psi^* _{n_r,m}), \label{cdcan} \\
\mathbf J^{\rm gau}&=& - \frac{e}{\mu} \mathbf A \psi^* _{n_r,m} \psi _{n_r,m} \label{cdgau}
\end{eqnarray}
are probability current densities associated with the canonical momentum and gauge field respectively. They were named the canonical current and the gauge  current respectively \cite{Jap3}.

Substituting the eigenstates (\ref{nes}) and the symmetric gauge (\ref{sg1}) into the current densities (\ref{cdcan}) and (\ref{cdgau}), we get
\begin{eqnarray}
\mathbf J^{\rm can} &=& {\rm J}^{\rm can} {\mathbf e_\phi} = \frac{m \hbar}{\mu r} R_{n_r,m}^2 \mathbf e_\phi, \quad  \label{jcan}\\
\mathbf J^{\rm gau} &=& {\rm J}^{\rm gau} {\mathbf e_\phi} = \frac{|e|B r}{2 \mu} R_{n_r,m}^2 \mathbf e_\phi. \label{jgau}
\end{eqnarray}
and therefore
\begin{equation}
\mathbf J = {\rm J} \mathbf e_\phi=  \frac{\hbar}{\mu a^2} (\frac{am}{r} + \frac{r}{2a} ) R_{n_r,m}^2 \mathbf e_{\phi}. \label{tpcd}
\end{equation}
Eqs. (\ref{jcan}, \ref{jgau}) and (\ref{tpcd}) demonstrate that in the symmetric gauge (\ref{sg1}), both $\mathbf J^{\rm can}$ and $\mathbf J^{\rm gau}$, and therefore $\mathbf J$ only have azimuthal component.

It should be emphasized that although we choose the symmetric gauge throughout this paper, the probability current density (\ref{pcd}) is invariant under the  transformation
\begin{eqnarray}
\mathbf A &\to& \mathbf A ^\prime = \mathbf A + \nabla \chi , \no \\
\psi & \to& \psi ^\prime = \psi e^{-ie \chi/ \hbar} \label{gt}
\end{eqnarray}
with $\chi$ being an arbitrary differential function.

Obviously, $\mathbf J^{\rm gau}$ is always positive. Nevertheless, the directions of $\mathbf J^{\rm can}$ depends on the angular quantum number $m$. For states with $m \geq 0$, the two terms  inside the parentheses in (\ref{tpcd}) have the same sign, so ${\rm J}  \geq 0$ holds everywhere in space. Therefore, the probability flows in the counterclockwise direction. However, for states with $m<0$, $ \mathbf J^{\rm can}$  and $\mathbf J^{\rm gau}$ have opposite signs. Setting the expression in parentheses in Eq. (\ref{tpcd}) to zero, we find the critical radius
\begin{equation}
r_c = \sqrt {2|m|} a. \label{cr}
\end{equation}
at which the two contributions balance each other.


The critical radius thus partitions the plane into three regions with distinct flow characteristics. They are
\begin{eqnarray}
\bullet &r < r_c, & \frac{a m}{r} + \frac{r}{2 a} < 0, \ \   \rm J <0     \label{cg1}\\
\bullet	&r > r_c  & \frac{a m}{r} + \frac{r}{2 a} > 0, \ \   \rm J >0   \label{cg2}\\
\bullet &r = r_c  & \frac{a m}{r} + \frac{r}{2 a} = 0, \ \   \rm J =0.  \label{cg3}
\end{eqnarray}

This spatial structure is intrinsic to the Landau levels. It arises  from the competition between $\mathbf J^{\rm can}$ and $\mathbf J^{\rm gau}$.  In the inner region ($r<r_c$), $\mathbf J^{\rm can}$ dominates and the net current is clockwise; in the outer region ($r>r_c$), $\mathbf J^{\rm gau}$ dominates and the net current is counterclockwise. The critical radius $r_c$ provides a precise quantitative characterization of this structure, marking the spatial location where  $\mathbf J^{\rm can}$ and $\mathbf J^{\rm gau}$ reach exact balance. Since the probability current density  is invariant under the  gauge transformation (\ref{gt}), the critical radius and the intrinsic partition it defines are also gauge invariant.

\section{The cancellation mechanism}

In the preceding section, we derived the critical radius from the  competition between $\mathbf J^{\rm can}$ and $\mathbf J^{\rm gau}$.  This critical radius naturally partitions the plane into an inner region, where the flow is clockwise, and an outer region, where the flow is counterclockwise. We now use this partition to analyze the physical origin of the vanishing total current for negative angular momentum Landau eigenstates.

The total probability current is defined as the radial integral of the probability current density. The result is \cite{Li}:
\begin{eqnarray}
i &=&  i^{\rm can} +  i^{\rm gau} = \int_0 ^\infty  {\rm J} ^{\rm can} dr + \int _0 ^\infty  {\rm J}^{\rm gau} dr \no \\
 &=& \left \{
 	\begin{aligned}
		&\frac{|e|B}{4 \pi \mu}, & m=0, \\
		&\frac{|e|B}{4 \pi \mu}(1+ \frac{m}{|m|}), & m \neq 0.
        \end{aligned}
\right. \label{configuration}
\end{eqnarray}

From this expression we see  that for positive $m$, the total current  agrees with the classical picture (\ref{cp}). For $m=0$, the total current equals exactly half of the classical value. However, for negative $m$, the total current vanishes identically. The vanishing of the current for negative $m$ has been studied in the literature from a geometric perspective \cite{Li}. In that picture, the electron motion is regarded as a classical circular orbit. One introduces  the operator of the square of the distance of the orbit center to the  origin \cite{Li, Gri}
\begin{equation}
r_0^2 = x_0^2 + y_0^2 \label{r0}
\end{equation}
where
\begin{eqnarray}
x_0 &=& x + \frac{\pi_y}{eB}, \no \\
y_0 &=& y - \frac{\pi_x}{eB},
\end{eqnarray}
and an operator  for the cyclotron radius
\begin{equation}
R^2 = (x- x_0)^2 + (y - y_0)^2. \label{R}
\end{equation}
The operators  $r_0^2$ and $R^2$ commute with Hamiltonian (\ref{ha1}) and their expectation values over the common eigenstates (\ref{nes})  are \cite{Li, Gri}
\begin{eqnarray}
\langle r_0 ^2 \rangle &=& 2(n_r + |m| - m +1) a^2,  \label{pr0}  \\
\langle R ^2 \rangle &=& 2(n_r + |m| + m +1) a^2. \label{pR}
\end{eqnarray}

By comparing these two length scales, the geometric interpretation concludes that the total current vanishes if the classical orbit, as characterized by expectation values $\langle R^2 \rangle$, does not enclose the coordinate origin \cite{Li}. This picture is both intuitive and illuminating.

However, the introduction of the critical radius (\ref{cr}) may offer an alternative  way of understanding the same phenomenon. 
Using the critical radius, the total current can be split into contributions from the inner and outer regions
\begin{eqnarray}
i = \int_0 ^\infty {\rm J} \ dr = \Big(\int _0 ^{r_c} + \int_{r_c} ^\infty \Big) {\rm J} \ dr.
\end{eqnarray}
Considering Eqs. (\ref{cg1}, \ref{cg2}) and  (\ref{configuration}), one concludes that for any radial quantum number $n_r$ and any negative $m$, the result is
\begin{equation}
\int_0 ^{r_c} {\rm J} \ dr = -\int_{r_c} ^\infty  {\rm J} \ dr.
\end{equation}

It indicates that  the clockwise current in the inner region  and the counterclockwise current in the outer region are equal in magnitude and opposite in direction. Their radial integrals cancel exactly, yielding a vanishing total current. Therefore, the vanishing of the  current, not because there is no flow, but because the inward and outward circulations balance each other exactly.

This explanation is based on the intrinsic  structure of the probability current density, which is  gauge invariant. The critical radius $r_c$ is defined through the zero of this gauge invariant density, and the regional currents obtained by integrating up to and beyond the critical radius are therefore gauge invariant as well.
It is worth noting that the critical radius is intimately related to the geometric parameters introduced (\ref{r0}, \ref{R}). From the expressions for the expectation values (\ref{pr0}, \ref{pR}), one obtains the following relation
\begin{equation}
r_c ^2 = \langle r_0^2 \rangle -\langle R^2 \rangle .
\end{equation}

Geometrically, this means that the critical radius is the distance from the origin to the point of tangency of a line drawn from the origin to the classical orbit. This connection allows us to supplement the existing geometric picture with information about the direction of rotation: inside the tangent point, the electron has a clockwise tendency with respect to the origin; outside the tangent point, the tendency is counterclockwise. When integrated along the radial direction, these two tendencies cancel one another.



\section{Partitioned kinetic Angular Momentum and the competition Structure}

In the previous section, we saw that the vanishing total current for negative angular momentum states results from an exact cancellation between the clockwise flow in the inner region and the counterclockwise flow in the outer region. This cancellation, however, does not fully capture the character of the eigenstates described  by the negative quantum number. A natural question arises: how does the negative quantum number $m$ manifest itself in the actual rotational motion of the electron? 

To answer this question, we turn to the kinetic angular momentum, which describes the kinetic rotation of the electron. In the symmetric gauge (\ref{sg1}), it takes the form
\begin{equation}
L ^{\rm kin} = \mathbf r \times \bm \pi = -i \hbar \frac{\pa}{\pa \phi} + \frac{eB}{2} r^2.
\end{equation}
The expectation value of this operator in an arbitrary eigenstate $|n_r, m \rangle$ is given by the following integral
\begin{eqnarray}
\langle L ^{\rm kin} \rangle 
&=&\hbar \int _0 ^\infty (m  + \frac{ r^2}{2 a^2}) R^2 r \ dr \no \\
&=& \int_0 ^\infty  (2 \pi \mu r {\rm J}_\phi) r \ dr. \label{evka}
\end{eqnarray}
In terms of the quantum numbers, this expectation value takes a compact form
\begin{equation}
\langle L ^{\rm kin} \rangle = (2 n_r + |m| +m +1) \hbar. \label{km}
\end{equation}
According to this expression, one concludes that for any eigenstate, the expectation value of the kinetic angular momentum is always  positive. 
The algebra gives us the final answer, but it does not tell us how this positive net rotation emerges from an eigenstate labeled by a negative angular quantum number. In particular, the algebraic result (\ref{km}) alone offers no insight into the competition  process the electron undergoes in its actual rotation.

The critical radius may provide the missing perspective. To quantify the rotational contributions from the two distinct flow regions, it is
convenient  to define the partitioned kinetic angular momentum operators. We define the kinetic angular momentum operator for the inner region
\begin{eqnarray}
L_{\rm in}^{\rm kin} = L_z ^{\rm kin} \Theta (r_c -r),
\end{eqnarray}
and for the outer region
\begin{eqnarray}
L_{\rm out}^{\rm kin} = L_z ^{\rm kin} \Theta (r -r_c)
\end{eqnarray}
where $\Theta(x)$ is the Heaviside step function, defined by
\begin{equation}
\Theta (x) = \left \{ \begin{aligned}
1, \quad x>0,  \\
0, \quad x<0. \end{aligned}
\right.
\end{equation}
Clearly, the sum of the inner and outer contributions yields the total kinetic angular momentum, i.e.,
\begin{equation}
L^{\rm kin} = L_{\rm in}^{\rm kin} + L_{\rm out}^{\rm kin}.
\end{equation}

The expectation values of these partitioned operators in an arbitrary eigenstate are then given by the radial integrals restricted to the corresponding regions
\begin{eqnarray}
\langle L_{\rm in} ^{\rm kin} \rangle = \int _0 ^{r_c} ( 2 \pi r {\rm J})r \ dr,  \label{innerL}  \\
\langle L_{\rm out} ^{\rm kin} \rangle = \int _{r_c} ^\infty ( 2 \pi r {\rm J})r \ dr. \label{outerL}
\end{eqnarray}

The integrand is the product of the gauge invariant probability current density and the radial distance, thus, this partitioned definition is  gauge invariant. 

According to  Eqs. (\ref{cg1}) and  (\ref{cg2}),
we can read that the inner kinetic angular momentum is negative while the outer kinetic angular momentum is positive. 
The net result of this competition is a positive total kinetic angular momentum which is independent of the angular momentum quantum number for $m<0$. Therefore, it seems   that the negative angular momentum quantum number plays no role in the actual rotational motion of the electron. However, a detailed  analysis shows that it is not the case.

Since  two integrals (\ref{innerL}) and (\ref{outerL}) admit no analytical solutions, we study them numerically. Some  numerical results are listed in the Table I.
\begin{table}[h]
    \centering      
    \caption{Expectation values of partitioned kinetic angular momenta for several eigenstates with  $n_r = 0$  and  $m < 0$.  $\langle L_{\mathrm{in}}^{\mathrm{kin}}\rangle$  and $\langle L_{\mathrm{out}}^{\mathrm{kin}}\rangle$ as well as $L^{\rm kin}$ are all in units of  $\hbar$.} 
    \label{table1} 

    \begin{tabular}{|c|c|c|c|c|c|}
        \hline             
        $m$ & -1 & -5 &-10 &-50&-100 \\ 
        \hline             
        $\langle L^{\rm kin}_{\rm in} \rangle$ &-0.104&-0.493&-0.834&-2.354&-3.513 \\
        \hline
        $ \langle L^{\rm kin}_{\rm out} \rangle$ & 1.104&1.493&1.834&3.354&4.513 \\
        \hline
        $L^{\rm kin}$ &1&1&1&	1&1 \\
        \hline
        \end{tabular} \\
\vspace{0.3em}
\end{table}

The numerical results in the above table show that as the magnitude of $m$ increases, the strengths of both the inner and outer rotations grow synchronously, yet their difference remains a constant. 
This means that the contribution of negative $m$ does not disappear; rather, it manifests itself concretely in the simultaneous enhancement of both rotational components.


The numerical results in the  Table I also reveal an important fact: in a negative $m$ eigenstate, the electron  carries two rotational components in opposite directions. The inner component originates from the negative $m$ contribution in the phase gradient of the wave function, which drives a clockwise circulation. The outer forward rotation reflects the dominant role of the gauge field. The positive total kinetic angular momentum is the net outcome of the competition between these two components.

As a supplementary analysis, we  present the numerical results of the partitioned kinetic angular momenta for a fixed negative $m$  and increasing radial quantum number in Table II. This table shows that as the radial quantum number increases, the total kinetic angular momentum increases linearly. These results  coincide with the algebraic result (\ref{km}).
\begin{table}[h] 
    \centering      
    \caption{Expectation values of partitioned kinetic angular momenta for eigenstates with fixed $ m = -1$  and increasing radial quantum number  $n_r$, units of  $\hbar$. } 
    \label{table2} 

    \begin{tabular}{|c|c|c|c|c|c|}
        \hline             
        $n_r$ & 1 & 5 &10 &30&50 \\ 
        \hline             
        $L^{\rm kin}_{\rm in}$ &-0.127&-0.081&-0.063&-0.038&-0.030 \\
        \hline
        $L^{\rm kin}_{\rm out}$ & 3.127&11.081&21.063&61.038&101.030 \\
        \hline
        $L^{\rm kin}$ &3&11&21&	61&101 \\
        \hline
        \end{tabular} \\
\end{table}

The results of the above table display an interesting phenomenon: as the quantum number $n_r$ increase, the absolute value of the inner counter-rotation  decreases, while the outer forward rotation grows. In the following, we shall give  a physical  explanation.

The Eq. (\ref{evka}) can be written  in the form
\begin{equation}
\langle L^{\rm kin} \rangle = \hbar \int _0 ^\infty g(r) \ dr 
\end{equation}
where $g(r) = (mr + \frac{r^3}{2 a^2}) R^2_{n_r, m}$ can be factorized as  $g(r) = f_1(r) f_2 (r)$ with $f_1(r)$ and $f_2(r)$ being defined
\begin{eqnarray}
f_1 (r) &=& m r + \frac{r^3}{2 a^2},  \label{f2}    \\
f_2 (r) &=& R^2_{n_r, m} (r). \label{f3}
\end{eqnarray}
For fixed $m$, the factor $f_1(r)$ is independent of the radial quantum number. 
In the inner region $r < r_c$, $f_1(r)$ is a  negative curve that vanishes at $r = 0$ and  $r = r_c$. Therefore, the dependence on $n_r$ falls on the radial probability density $f_2(r)$. As the radial quantum number increases, the number of nodes in the radial wave function grows and the main peak of the probability density  shifts toward the origin, while several additional smaller probability peaks appear, distributed in both the inner and outer regions. All of these changes are subjected to the rigid constraint of the normalization condition, i.e.,
\begin{equation}
\int _0 ^\infty r R^2_{n_r, m} (r)  \ dr =1.
\end{equation}


\begin{figure}[!htb]
\centering
\makebox[\linewidth]{\includegraphics[width=1.0\linewidth]{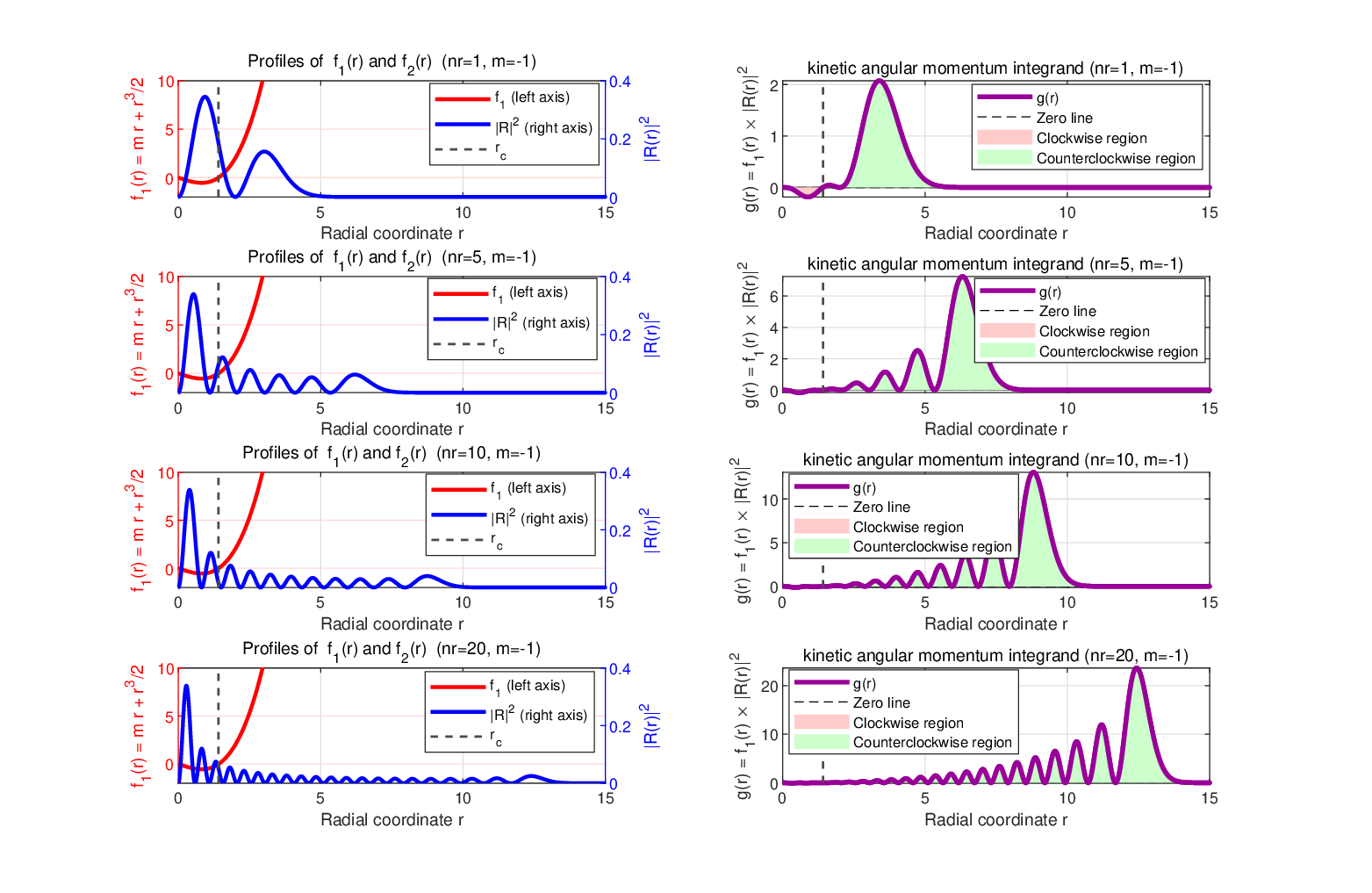}}
\caption{The factors $f_1(r)$, $f_2(r)$ and partitioned kinetic angular momentum integrand for Landau eigenstates with $m=-1$ and varying radial quantum numbers $n_r$  (setting magnetic length $a=1$).
Left panels display the  factor $f_1(r) = m r + \half r^3$ (blue curve, left axis) and the radial probability density $f_2(r) = R^2_{n_r, m}$  (red curve, right axis). The blue dashed line marks the critical radius $r_c = \sqrt {2|m|}$. Right panels show the kinetic angular momentum integrand $g(r) = f_1(r) f_2(r)$.  The pink-shaded regions indicate negative (clockwise, CW) contributions, while the green-shaded regions indicate positive (counterclockwise, CCW) contributions. As $n_r$ increases, the radial probability density shifts outward beyond $r_c$, causing the probability weight in the inner region to decrease and the outer CCW contribution to dominate, which accounts for the linear growth of the net kinetic angular momentum $\langle L^{\rm kin} \rangle =(2 n_r +1) \hbar$.}
\end{figure}






This structure has a direct impact on the kinetic angular momentum in the inner region: since $f_1 (r)$ does not change with respect to $n_r$ and takes very small values near the origin, the inward shift of the main probability peak places the largest share of the probability density in a region where $f_1(r)$ is smaller. The product $f_1(r) f_2(r)$ in the inner region is thereby gradually suppressed, and the negative integrated contribution $\langle L_{\rm in}^{\rm kin} \rangle$ decreases.
In the outer region $f_1(r)$ grows as $r^3$, and the additional small probability peaks act on $f_1(r)$, giving extra probability weight to the outer region. 
The inner region provides only a negative contribution, and its magnitude becomes negligible as $n_r$ grows. The net result after cancellation is the clear linear growth $\langle L^{\rm kin} \rangle  = (2n_r+1)\hbar$.

This behavior should be compared with the situation in Table 1. When $n_r$ is fixed and $|m|$ increases, the probability density neither shifts nor develops new small peaks in the outer region. Instead, the critical radius $r_c = \sqrt {2 |m|} a$ expands, and the factor $f_1(r)$ undergoes a global rescaling. Consequently, the inner and outer contributions grow together. 

Thus, the factorization into $f_1(r)$ and $f_2(r)$ provides a unified framework for understanding both tables. In the large $n_r$ limit, the main peak of the probability density moves far from $r_c$ and approaches the origin, the probability weight in the inner region tends to zero, and the kinetic angular momentum becomes entirely dominated by counterclockwise rotation in the outer region. In this limit, 
although the intrinsic bidirectional flow structure persists in the exact quantum mechanical sense, nearly all of the electron's probability distribution resides in the outer region, and the total rotation appears as a unidirectional counterclockwise cyclotron motion. This behavior is consistent with the correspondence principle: in the limit of large radial quantum numbers, the effective observable behavior of the quantum state approaches the classical picture \cite{Bohr}.

\section{Conclusions and Remarks}
In this paper, we have analyzed the probability current density of Landau eigenstates in a uniform magnetic field. For states with a negative angular quantum number  $m$, the competition between the canonical-momentum-related  and gauge field contributions gives rise to a critical radius  $r_c = \sqrt{2|m|}a$, which partitions the plane into an inner clockwise flow and an outer counterclockwise flow. Based on this structure, we obtain two main results. First, the vanishing total current for  $m<0$  is not an absence of flow, but an exact radial cancellation between the opposing currents on the two sides of  $r_c$. Second, by partitioning the kinetic angular momentum with respect to  $r_c$, we find that the negative quantum number  $m$  manifests itself as a simultaneous increase of both the inner and outer rotational strengths, while their difference, i.e., the net kinetic angular momentum, remains fixed.

As a by-product of our previous studies, we hope that this bidirectional competition may also present an alternative interpretation  for the internal dynamics of different degenerate states within the same Landau level. Since the energy is proportional to the net kinetic angular momentum,  $E_{n_r,m} = \frac{|e|B}{2\mu}\langle L^{\mathrm{kin}}\rangle$, different negative $m$  states with the same  $n_r$  share the same net kinetic angular momentum and hence the same energy, despite having completely different internal flow compositions (as seen in Table I). The degeneracy thus reflects the fact that infinitely many distinct internal competition patterns yield the same net rotation.

The bidirectional flow structure is an exact quantum property. In the limit of large radial quantum numbers, the probability density is localized far beyond  $r_c$, making the inner region negligible and recovering the classical unidirectional cyclotron motion. Therefore, this work offers a new perspective on the quantum-classical correspondence and on the internal dynamics of Landau levels.

\section*{Funding}
Funding information - not applicable.

\section*{Author Contributions}
J. M. Wang: Methodology, Formal analysis, Conceptualization; Y. Z. Gao and D. L. Cun:  Analytical and Numerical calculations;  J. Jing: Writing
original draft, Validation, Formal analysis.

\section*{Data availability}
No datasets were generated or analysed during the current study.

\section*{Declaration of competing interest}
The authors declare that they have no known competing financial interests or personal relationships that could have appeared to influence the work reported in this paper.

\end{document}